\documentclass[10pt]{article}
\usepackage{amssymb}
\usepackage{amsmath}
\usepackage{amsthm}
\usepackage{latexsym}
\usepackage[dvips]{epsfig}
\usepackage{mathrsfs}
\usepackage{eufrak}

\theoremstyle{plain}

\newtheorem{lemma}{Lemma}

\newtheorem{assumption}{Assumption}

\newtheorem{corollary}{Corollary}
\newtheorem*{main}{Theorem}
\newtheorem*{definition}{Definition}

 \font\tenscr=rsfs10 scaled1100
\font\sevenscr=rsfs7 
\font\fivescr=rsfs5 
\skewchar\tenscr='177 \skewchar\sevenscr='177 \skewchar\fivescr='177
\newfam\scrfam
\textfont\scrfam=\tenscr \scriptfont\scrfam=\sevenscr
\scriptscriptfont\scrfam=\fivescr

\def\scriL{{\fam\scrfam L}}

\newcommand{\pd}[2]{\frac{\partial #1}{\partial #2}}

\font\tenscr=rsfs10 scaled1100
\font\sevenscr=rsfs7 
\font\fivescr=rsfs5 
\skewchar\tenscr='177
\skewchar\sevenscr='177
\skewchar\fivescr='177
\newfam\scrfam
\textfont\scrfam=\tenscr
\scriptfont\scrfam=\sevenscr
\scriptscriptfont\scrfam=\fivescr

\begin{document}

\title{\textbf{Static elastic shells in Einsteinian and Newtonian gravity}}

\author{Christiane Maria Losert-Valiente Kroon \thanks{E-mail address:
 {\tt christianelosert@gmx.at}} \\
 Institut f\"ur Theoretische Physik,\\ Universit\"at Wien,\\
 Boltzmanngasse 5, A-1090 Wien,\\ Austria.}

\maketitle

\begin{abstract}
We study the behaviour of a specific system of relativistic
elasticity in its own gravitational field: a static, spherically
symmetric shell whose wall is of arbitrary thickness consisting of
hyperelastic material. We give the system of field equations and
boundary conditions within the framework of the Einsteinian theory
of gravity. Furthermore, we analize the situation in the Newtonian
theory of gravity and obtain an existence result valid for small
gravitational constants and pointwise stability by using the
implicit function theorem. If one replaces the elastic material with
a fluid, one finds that stable states can not exist.

\end{abstract}

 \section{Introduction}

Let us take any static spherically symmetric shell consisting of an
elastic material. Picture the shell to be a sphere from which a
sphere of smaller radius is cut out around the center of mass of the
bigger sphere and replace the resulting hole by vacuum.  The elastic
material is assumed to be isotropic and homogeneous. At first,
ignore the own gravitational field of the elastic shell. When
``switching on" the shell's self-gravitating field one will observe
a deformation of the static elastic shell. In order to describe this
process one can make use of either of the following two pictures:
the \emph{spatial description} (\emph{Euler picture}) or the
\emph{material description} (\emph{Lagrange picture}). In any of the
pictures the undeformed body, $\Bbb{B}$, is represented by a
three-dimensional differentiable manifold that is endowed with a
flat metric (\emph{body metric}), $G_{AB}$ --- $A,B=1,2,3$. The
spacetime, $\Bbb{M}$, a four-dimensional manifold, is equipped with
the \emph{spacetime metric} $g_{\alpha \beta}$ --- $\alpha, \beta
=0,1,2,3$. We assume that there exists a \emph{natural state} or
\emph{relaxed state} of the static elastic shell, i.e. one should
understand this state as to be strain- and hence stressfree. In the
spatial description we choose local coordinates $y^\alpha$ on
$\Bbb{M}$ and let the motion of the body be given by a surjective
mapping $z$ --- the so-called deformation map --- from the spacetime
onto the body so that $Y^A=z^A(y^\alpha)$ are coordinates on
$\Bbb{B}$. The field equations describing the behaviour of the
static elastic shell viewed from the point of the spatial
description are the Einstein field equations with elastic matter
source on $\Bbb{M}$. For the material description we have to assume
that the spacetime has a foliation consisting of spacelike
hypersurfaces. The motion of the body is given by the deformation
map $f$ from the undeformed body onto a submanifold in spacetime,
namely $f(\Bbb{B}) \times \Bbb{R}$. The tensor field
$g_{\alpha\beta}$ solves the Einstein vacuum field equations on
$\Bbb{M}/f(\Bbb{B})\times\Bbb{R}$ and the Einstein field equations
with elastic matter source composed with the deformation map $f$ on
$f(\Bbb{B})\times\Bbb{R}$. For the static case, the deformation map
$z^A$ of the spatial description is the inverse to the deformation
map $f^a$ --- $a=1,2,3$ --- of the material description.
\\
From the experimental point of view it is clear that stable states
for the above described configuration exist for elastic materials
but do not exist in the case of fluids. It would be of interest to
derive this from the theory of elastomechanics. In \cite{BeiSch;02}
an existence proof for general static, self-gravitating elastic
bodies in the Newtonian theory of gravity was given which includes
the static elastic shell as a special case. Here we follow a
different approach to derive an existence result and, in addition,
explicitly write down the field equations and boundary conditions
for the static elastic shell in both the Einsteinian and Newtonian
theory.\\
In the following section we will introduce some important quantities
of elasticity that will simplify the tackling of our setting. We
will make some crucial assumptions on the material under
consideration. We obtain a nice expression for the energy-momentum
tensor in terms of typical objects of elastomechanics and give the
Einstein field equations and boundary conditions in the material
description. In section 3 we consider our system of equations in the
Newtonian limit. Still, the Newtonian field equation and boundary
conditions cannot be solved explicitly for general elastic matter.
Therefore, section 4 is concerned with an analytical approach to
prove that for small gravitational constants and in the case of
pointwise stability stable states of the static elastic shell exist.
In the appendix, we solve the linearised system of Newtonian
equations and illustrate the behaviour of the static elastic shell
by means of three concrete examples.

\section{Field equations and boundary conditions in the Einsteinian theory of gravity}

\subsection{Energy-momentum tensor}

We derive the field equations, namely the Einstein equations, from a
Lagrangian principle in the spatial description. We aim to present
the energy-momentum tensor using quantities within the framework of
the theory of elastomechanics. Let us start with the general
definition of the energy-momentum tensor:
\begin{definition}
The energy-momentum tensor is defined by
\begin{equation*}
t_{\alpha\beta}:=2\frac{\delta\mathcal{L}}{\delta
g^{\alpha\beta}}-\mathcal{L}g_{\alpha\beta},
\end{equation*}
\noindent where $\mathcal{L}$ is the Lagrangian density.
\end{definition} \noindent
In our case, the Lagrangian density is given by
\begin{equation*}
\mathcal{L}=n(\rho_0c^2+w),
\end{equation*}
\noindent where $n$ denotes the particle number density. The first
term in brackets gives the energy density of the relaxed state and
the second term, $w$, is called the \emph{stored energy function
density}. It is a very important quantity of elastomechanics and we
will enlighten this in the few following paragraphs. In general, the
stored energy function depends on the deformation map, its
derivatives with respect to the coordinates on spacetime and the
coordinates on spacetime:
\begin{equation*}
w=w[z^A,\pd{z^B}{y^\alpha},y^\beta].
\end{equation*}
\noindent One can rewrite the energy-momentum tensor using the
expression for the Lagrangian density such that one gets:
\begin{equation*}
t_{\alpha \beta}=\mathcal{L}u_{\alpha}u_{\beta}+ S_{\alpha \beta},
\end{equation*}\noindent
where $u^{\alpha}$ is the \emph{4-velocity} --- a timelike, future
pointing vector field fulfilling $u^{\alpha}\pd{z^A}{y^\alpha}=0$
and $g_{\alpha \beta}u^\alpha u^\beta=-1$. The term $S_{\alpha
\beta}$ is called the \emph{stress tensor} and satisfies $S_{\alpha
\beta}u^{\beta}=0$. The stored energy function determines the
terminology of elastic materials, for example via the following
definition:
\begin{definition}
If there exists a stored energy function such that the stress tensor
takes the form
\begin{equation*}
S_{\alpha \beta}=2n\pd{w}{g^{\alpha \beta}},
\end{equation*}
then the material is said to be hyperelastic.
\end{definition}
\noindent All information about the specific classes of materials
considered is contained in the stored energy function. Reminding of
the form of the Lagrangian density one finds that we are actually
dealing with hyperelastic material. Next, we want to introduce
another important object of elasticity that is known as the
\emph{Cauchy-Green strain tensor}. Its name is justified when
considering the fact that in spherical symmetry the $(R,R)$
component of the strain tensor $(H^A{}_B)$ measures the compression,
$(H^R{}_R)<0$, as well as the stretching, $(H^R{}_R)>0$, of a
material. In general, we have:
\begin{definition} The strain tensor is defined as
\begin{equation*}
H^{AB}:=g^{\alpha\beta}\pd{z^A}{y^{\alpha}}\pd{z^B}{y^{\beta}},
\end{equation*}
\noindent and
\begin{equation*}
H^A{}_B=H^{AC}G_{CB}.
\end{equation*}
\end{definition}
\noindent
Covariance of $w$ under spatial diffeomorphism gives us
\begin{equation*}
w=w[z^A,H^{BC}].
\end{equation*}
\noindent Let us now assume that the stored energy function depends
only on some invariants,$J_i$, of the strain tensor $H^{A}{}_{B}$:
\begin{equation*}
w=w[J_i].
\end{equation*}
\noindent We make use of a certain choice of invariants\footnote{For
the verification of the fact that the square root of the determinant
of the strain tensor is equal to the inverse number density we refer
to \cite{BeiSch;03}.} of the strain tensor, $H^A{}_B$, namely:
\begin{eqnarray*}
&&J_1=tr(H^A{}_B),\\
&&J_2=\frac{1}{2}[tr(H^A{}_B)^2-tr(H^A{}_CH^C{}_B)],\\
&&J_3=\det(H^A{}_B)=n^{-2}.
\end{eqnarray*}
 \noindent At \emph{natural points}
$\big(y^\alpha|_0,Y^A|_0,z^B|_0\big)$ the strain tensor $H^{AB}$
takes the following expression
\begin{equation*}
H^{AB}|_0:=H^{AB}\big(\pd{z^C}{y^\alpha}|_0,g^{\mu
\nu}(y^\beta|_0)\big)=G^{AB}(Y^C|_0),
\end{equation*}
\noindent in other words, the state of the static elastic shell at
natural points is strainfree. In the relaxed state the chosen
invariants of the strain tensor reduce to the following constants
 \begin{equation*}
J_1|_0=J_2|_0=3, \quad J_3|_0=1.
 \end{equation*}\noindent
The stored energy function has to vanish and have a minimum in a
locally relaxed state of the matter. From the expansion of the
stored energy function at natural points we see that for isotropic,
homogeneous materials there are constants $\lambda$ and $\mu$ ---
the \emph{Lam\'e constants} --- such that
\begin{equation*}
\frac{\partial^2w}{\partial H^{AB}\partial
H^{CD}}=\frac{\lambda}{4\rho_0}H_{AB}H_{CD}+\frac{\mu}{2\rho_0}H_{C(A}H_{B)D}.
\end{equation*}
\noindent We reduce our analysis to isotropic and homogeneous
material. Furthermore, we require our system to fulfill pointwise
stability. In the case of isotropic, homogeneous elastic material
--- see, for example, \cite{MarHug;83} --- this condition can be expressed as
\begin{equation*}
\mu>0, \quad 3\lambda+2\mu>0.
\end{equation*}
\noindent Now, we will use all of the aforementioned assumptions and
definitions to rewrite the energy-momentum tensor in terms of
objects typical of the theory of elasticity. Due to the form of the
boundary conditions it is more convenient to turn to the material
description. Therefore, we derive the energy-momentum tensor in the
material description, $T^{\alpha\beta}$, from the one of the spatial
description, $t_{\alpha\beta}$, and then compose it with the
deformation map $f^\alpha$. The energy-momentum tensor in the
material description reads
\begin{equation*}
T^{\alpha\beta}=(J_3)^{-\frac{1}{2}}\bigg[(\rho_0c^2+w)u^\alpha
u^\beta-2\big(\pd{w}{J_1}G^{AB}+(J_1G^{AB}-H^{AB})\pd{w}{J_2}+J_3\tilde{H}^{AB}
\pd{w}{J_3}\big)\pd{f^{\alpha}}{X^A}\pd{f^{\beta}}{X^B}\bigg],
\end{equation*}
\noindent where the quantities $J_i$
---$i=1,2,3$---,$w$,$u^{\alpha}$ and $H_{AB}$ are to be understood as
 the analogous objects in the material description to the ones we treated
 earlier in this section in the spatial description. The symbol
$\tilde{H}^{AB}$ is meant to be the inverse of $H_{AB}$. Note that
$\tilde{H}^{AB}$ is different to $H^{AB}$. The latter results of
rising the indices of the strain tensor in the material description
$H_{AB}$ with $G^{AB}$.

\subsection{Field equations}

Having obtained a nice expression for the energy-momentum tensor in
the material description we now turn our attention towards the
equations describing the process of deformation of the static
elastic shell taking into account the influence if the shell's
self-gravitating field. The required field equations in the material
description are the Einstein vacuum equations outside the deformed
body and the Einstein field equations with an elastic matter source
composed with the deformation map on the deformed body. That is, on
$\Bbb{M}/f(\Bbb{B})\times \Bbb{R}$ we have
\begin{equation*}
G^{\alpha\beta}(X^A)=0,
\end{equation*}
\noindent and on $f(\Bbb{B})\times\Bbb{R}$
\begin{equation*}
G^{\alpha\beta}(f^\gamma(X^A))=\kappa
T^{\alpha\beta}(f^\gamma(X^A)),
\end{equation*}
\noindent where $G^{\alpha \beta}$ is the Einstein curvature tensor,
$\kappa:= \frac{8 \pi \mathcal{G}}{c^{4}}$, $\mathcal{G}$ is the
gravitational constant, $c$ is the speed of light and $T^{\alpha\beta}$ is the energy-momentum tensor.\\ \\
We choose coordinates $X^A=(R,\theta,\phi)$ on the body $\Bbb{B}$
and introduce on spacetime $\Bbb{M}$ the coordinates
$x^\alpha=(ct,x^a)$ where $x^a=f^a(X^A)=(r=F(R), \vartheta=\theta,
\varphi=\phi)$ --- $F(R)$ being a monotone function. The exterior
Schwarzschild metric\footnote{This makes sense since we will require
asymptotic flatness and we, therefore, take into account the
Birkhoff's theorem --- see \cite{Wal;84}.} is matched to the outer
boundary of the shell, that is where $R=R_o$ is the outer radius.
The hollow region in the centre of the shell (where $R\leq R_i$,
$R_i$ is the inner radius) is described by a flat metric. The body
metric reads
\begin{equation*}
G_{AB}=diag(1,R^2,R^2\sin^2\theta).
\end{equation*}
\noindent We divide the spacetime in three regions corresponding,
respectively, to the hollow centre, the deformed body itself and the
Schwarzschildean exterior of the shell:
\begin{eqnarray*}
&&\stackrel{1}{g}{}_{\alpha \beta}=diag(-C,1,F^{2}(R),F^{2}(R)\sin^{2}{\theta}),\\
&&\stackrel{2}{g}{}_{\alpha \beta}=diag(-A[F(R)],B[F(R)],F^{2}(R),F^{2}(R)\sin^{2}{\theta}),\\
&&\stackrel{3}{g}{}_{\alpha
\beta}=diag(-\left(1-\frac{2\mathcal{G}M}{c^{2}F(R)}\right),\left(1-\frac{2\mathcal{G}M}{c^{2}F(R)}\right)^{-1},F^{2}(R),F^{2}(R)\sin^{2}{\theta}),
\end{eqnarray*}
\noindent where $C$ is a positive constant and $M$ is the central
mass.\\ \\
We require asymptotic flatness, that is
\begin{eqnarray*}
&&g_{tt}[F(R)] \longrightarrow -1 \ \ \ \text{as} \ F\rightarrow \infty, \\
&&g_{rr}[F(R)] \longrightarrow 1 \ \ \ \text{as} \ F\rightarrow
\infty.
\end{eqnarray*}
\noindent In a slight abuse of notion we write
$T_{\alpha}{}^{\rho}=\stackrel{2}{g}_{\alpha \beta}T^{\beta \rho}$,
and get
\begin{eqnarray*}
\nonumber
&&T_{t}{}^{t}=-\left[\frac{dF}{dR}\frac{\sqrt{B}F^{2}}{R^{2}}\right]^{-1}(\rho_0c^2+w),\\
\nonumber
&&T_{r}{}^{r}=-2\frac{R^{2}}{F^{2}}\frac{dF}{dR}\sqrt{B}\left(\pd{w}{J_{1}}+2\frac{F^{2}}{R^{2}}\pd{w}{J_{2}}+\frac{F^{4}}{R^{4}}\pd{w}{J_{3}}\right),\\
\nonumber
&&T_{\vartheta}{}^{\vartheta}=T_{\varphi}{}^{\varphi}\\
\nonumber
&&\phantom{T_{\vartheta}{}^{\vartheta}}=-2(\frac{dF}{dR}\sqrt{B})^{-1}\left[\pd{w}{J_{1}}+
\left(\big(\frac{dF}{dR}\big)^{2}B+\frac{F^{2}}{R^{2}}\right)\pd{w}{J_{2}}+\big(\frac{dF}{dR}\big)^{2}B\frac{F^{2}}{R^{2}}\pd{w}{J_{3}}\right].
\end{eqnarray*}
\noindent Keeping these expressions in mind, and after having
computed the Einstein curvature tensor we write the Einstein
equations on $f(\Bbb{B})\times \Bbb{R}$ in the following way:

\begin{eqnarray}
\label{1st einstein equation}
&&\kappa T_{t}{}^{t}=-\frac{\frac{\partial B}{\partial F}}{B^{2}F}-\frac{1}{F^{2}}+\frac{1}{BF^{2}},\\
\label{2nd einstein equation}
&&\kappa T_{r}{}^{r}=\frac{\frac{\partial A}{\partial F}}{ABF}-\frac{1}{F^{2}}+\frac{1}{BF^{2}},\\
\nonumber &&\kappa T_{\vartheta}{}^{\vartheta}=\kappa
T_{\varphi}{}^{\varphi}=-\frac{\frac{\partial B}{\partial
F}}{2B^{2}F}+\frac{\frac{\partial A}{\partial F}}{2ABF}
+\frac{\frac{\partial^2A}{\partial
F^2}}{2AB}-\frac{\big(\frac{\partial A}{\partial
F}\big)^{2}}{4A^{2}B}-
\frac{\frac{\partial A}{\partial F}\frac{\partial B}{\partial F}}{4AB^{2}}.\\
\label{3rd einstein equation}
\end{eqnarray}
\noindent If the first and second Einstein equations --- \eqref{1st
einstein equation} and \eqref{2nd einstein equation} --- hold then
one can, instead of equation \eqref{3rd einstein equation}, consider
--- see, for example, \cite{Wal;84} --- the only part of the conservation
law that is not fulfilled identically, namely:
\begin{equation*}
\nabla_{\alpha}T_r{}^\alpha=0,
\end{equation*}
\noindent which can be written as
\begin{eqnarray}
\label{conservation law F(R)} \nonumber &&\big[\frac{\frac{\partial
A}{\partial F}\big(\frac{dF}{dR}\big)^{2}B}{A}+2\frac{\partial
B}{\partial F}\big(\frac{dF}{dR}\big)^{2}+
2B\frac{d^2F}{dR^2}-\frac{\frac{\partial B}{\partial
F}\frac{dF}{dR}}{2B}+\frac{4B\frac{dF}{dR}}{R}-\frac{4F}
{R^{2}}\big]\pd{w}{J_{1}}+2B\frac{dF}{dR}\frac{\partial^{2}{w}}{\partial{J_{1}}\partial{R}}+\\
\nonumber &&\big[2\frac{F^{2}}{R^{2}}\big(\frac{\frac{\partial
A}{\partial F}\big(\frac{dF}{dR}\big)^{2}B}{A}+2\frac{\partial B}
{\partial
F}\big(\frac{dF}{dR}\big)^{2}+2B\frac{d^2F}{dR^2}-\frac{\frac{\partial
B}{\partial F}\frac{dF}{dR}}{2B}\big)+
\frac{4\big(\frac{dF}{dR}\big)^{2}BF}{R^{2}}-\frac{4F^{3}}{R^{4}}\big]\pd{w}{J_{2}}+\\
\nonumber
&&\frac{4B\frac{dF}{dR}F^{2}}{R^{2}}\frac{\partial^{2}{w}}{\partial{J_{2}}\partial{R}}+\\
\nonumber &&\big[\frac{F^{4}}{R^{4}}\big(\frac{\frac{\partial
A}{\partial F}\big(\frac{dF}{dR}\big)^{2}B}{A}+ 2\frac{\partial
B}{\partial
F}\big(\frac{dF}{dR}\big)^{2}+2B\frac{d^2F}{dR^2}-\frac{\frac{\partial
B}{\partial F}
\frac{dF}{dR}}{2B}\big)+\frac{4\big(\frac{dF}{dR}\big)^{2}BF^{3}}{R^{4}}-\frac{4B\frac{dF}{dR}F^{4}}{R^{5}}\big]
\pd{w}{J_{3}}+\\
&&\frac{2B\frac{dF}{dR}F^{4}}{R^{4}}\frac{\partial^{2}{w}}{\partial{J_{3}}\partial{R}}-(\rho_{0}c^{2}+w)
\frac{\frac{\partial A}{\partial F}}{2A}=0.
\end{eqnarray}
\noindent For convenience, we choose equation \eqref{conservation
law F(R)} to substitute for equation \eqref{3rd einstein equation}
in our further investigations and consider equation
\eqref{conservation law F(R)} together with the first and second
Einstein field equation, \eqref{1st einstein equation} and
\eqref{2nd einstein equation}, the demanded system of field
equations.

\subsection{Boundary conditions}
Clearly, additional equations will have to be fulfilled at the inner
and outer boundaries of the deformed body in order to satisfy the
standard matching conditions --- see, for example, \cite{MarSen;88}.
The fundamental boundary conditions of our system have to be derived
from the condition that the first fundamental form,
$g_{\alpha\beta}$, and the second fundamental form,
$K_{\alpha\beta}$, of the three metrics $\stackrel{1}{g}_{ab}$,
$\stackrel{2}{g}_{ab}$, $\stackrel{3}{g}_{ab}$ $(a,b=1,2,3$ spatial
indices) have to coincide at the inner and outer boundaries of the
static elastic shell. We consider the hypersurface
$\mathcal{H}:F(R)=const$ and the normal vector field of
$\mathcal{H}$, $n^\alpha$, with
\begin{equation*}
n^\alpha \pd{}{x^\alpha}=(g^{rr})^\frac{1}{2}\pd{}{r}.
\end{equation*}\noindent
The induced metric $h_{\alpha \beta}$ on $\mathcal{H}$ is
Lorentzianlike, $n_{\alpha}$ being spatial.\\
Therefore, we have
\begin{eqnarray*}
&&h_{\alpha\beta}dx^\alpha
dx^\beta=g_{tt}c^2dt^2+F(R)^2(d\theta^2+\sin^2\theta
d\phi^2),\\
&&K_{\alpha\beta}=\frac{1}{2}\scriL_nh_{\alpha\beta}=\frac{1}{2}(g^{rr})^{\frac{1}{2}}
\big(\frac{dF}{dR}\big)^{-1}\pd{}{R}h_{\alpha\beta},
\end{eqnarray*}
\noindent where $\scriL$ denotes the Lie derivative. Considering the
hypersurface $\mathcal{H}_i:F(R)=F(R_i)$ it follows that
\begin{eqnarray}
\label{inner boundary}
&&A[F(R_i)]=constant,\\
&&B[F(R_i)]=1.
\end{eqnarray}
\noindent The boundary conditions derived from the matching
conditions at the hypersurface $\mathcal{H}_o:F(R)=F(R_o)$ read
\begin{eqnarray}
\label{outer boundary}
&&A[F(R_o)]=1-\frac{2\mathcal{G}M}{c^2F(R_o)},\\
&&B[F(R_o)]=A[F(R_o)]^{-1}.
\end{eqnarray}
\noindent As a consequence of the 3+1 decomposition of the manifold
we also have to look on the constraint equations:
\begin{eqnarray}
\label{Hamilton constraint}
&&{}^{(3)}R[h]+(K^{a}{}_{a})^{2}-K_{ab}K^{ab}=2T_{ab}n^{a}n^{b},\\
\label{momentum constraint}
&&D^{b}(K_{ab}-K^{c}{}_{c}h_{ab})=(\phi^{*})^{d}{}_{a}(T_{db}n^{b}),
\end{eqnarray}
\noindent where ${}^{(3)}R[h]$ is the Ricci scalar with respect to
the three-dimensional metric $h$, $D^{d}M_{ab}:=h^{d}{}_{d
'}h_{a}{}^{a
'}h_{b}{}^{b '}\nabla^{d'}M_{a ' b '}$ is the projective derivative, and
$(\phi^{*})^{a}{}_{b}$ is the pull-back of the embedding.\\ \\
The momentum constraint equations \eqref{momentum constraint} are
fulfilled identically, whereas the Hamilton constraint equation
\eqref{Hamilton constraint} leads to the following boundary
condition:
\begin{equation*}
T_{rr}n^{r}|_{\partial(f(\Bbb{B}))}=0,
\end{equation*}
\noindent which is equivalent to
\begin{equation}
\label{boundary condition}
\frac{\sqrt{B}F^2}{R^2}T_{rr}n^R|_{\partial\Bbb{B}}=0,
\end{equation}
\noindent where $n^R$ is the normal vector of the hypersurface
$R=constant$.
\\ Henceforth, we are able to give the entire system of field
equations ---\eqref{1st einstein equation}, \eqref{2nd einstein
equation} and \eqref{conservation law F(R)}--- and boundary
conditions ---\eqref{inner boundary}, \eqref{outer boundary} and
\eqref{boundary condition}--- of the static elastic shell in its own
gravitational field within the framework of the Einsteinian theory
of gravity. Let us now investigate the situation in the Newtonian
theory of gravity.

\section{Newtonian limit}
Our aim  is to derive the system of the Newtonian equations from the
results we obtained in the latter section. From the first Einstein
equation \eqref{1st einstein equation} we derive the following form
of the field $B[F(R)]$, namely:
\begin{equation*}
B[F(R)]=\bigg(1-\frac{2\mathcal{G}m[F(R)]}{c^2F(R)}\bigg)^{-1},
\end{equation*}
\noindent where
\begin{equation*}
m[F(R)]=\frac{4\pi}{c^2}\int_{R_i}^R
F^2(\bar{R})\big(\rho_0c^2+w[F(\bar{R})]\big)d\bar{R}.
\end{equation*}
\noindent We insert the latter expression for the unknown field
$B[F(R)]$ as well as the following form of the unknown field
$A[F(R)]$, namely
\begin{equation*}
A[F(R)]=c^2 \exp \frac{2U[F(R)]}{c^2},
\end{equation*}
\noindent --- where $U$ can be viewed as a potential --- into the
system of field equations \eqref{1st einstein equation}, \eqref{2nd
einstein equation}, \eqref{conservation law F(R)} and boundary
conditions \eqref{boundary condition}. In order to obtain the system
of field equations and boundary conditions in the Newtonian theory
of gravity we take the limit $c \longrightarrow \infty$. In this
Newtonian limit the first two Einstein equations \eqref{1st einstein
equation} and \eqref{2nd einstein equation} reduce to the Poisson
equation
\begin{equation}
\label{Poisson equation}
\triangle U=4\pi \mathcal{G}(\rho)_{newt},
\end{equation}
\noindent and the conservation law \eqref{conservation law F(R)} in
the Newtonian limit gives the load equation
\begin{equation}
\label{load equation}
\frac{2}{F}\big((T_{r}{}^{r})_{newt}-(T_{\vartheta}{}^{\vartheta})_{newt}\big)
+\big(\frac{dF}{dR}\big)^{-1}(\pd{}{R}T_{r}{}^{r})_{newt}=-\frac{\partial
U}{\partial F}(\rho)_{newt},
\end{equation}
\noindent where
\begin{eqnarray*}
&&(T_{r}{}^{r})_{newt}=-2\frac{R^{2}}{F^{2}}\big(\frac{dF}{dR}\big)^{-1}
\left[\pd{w}{J_{1}}+2\frac{F^{2}}{R^{2}}\pd{w}{J_{2}}+\frac{F^{4}}{R^{4}}\pd{w}{J_{3}}\right],\\
&&(T_{\vartheta}{}^{\vartheta})_{newt}=(T_{\varphi}{}^{\varphi})_{newt}=-2\big(\frac{dF}{dR}\big)^{-1}
\left[\pd{w}{J_{1}}+\left(\big(\frac{dF}{dR}\big)^{2}+\frac{F^{2}}{R^{2}}\right)\pd{w}{J_{2}}+
\frac{F^{2}}{R^{2}}\big(\frac{dF}{dR}\big)^{2}\pd{w}{J_{3}}\right]\\
&&(\rho)_{newt}=\frac{R^{2}}{F^{2}}\big(\frac{dF}{dR}\big)\rho_{0}.
\end{eqnarray*}
\noindent The remaining boundary conditions in the Newtonian limit
are
\begin{equation}
\label{boundary Newton}
\frac{F^{2}}{R^{2}}(T_{r}{}^{r})_{newt}\mid_{\partial{\Bbb{B}}}=0.
\end{equation}
\noindent Equations \eqref{Poisson equation}, \eqref{load equation}
and \eqref{boundary Newton} form the complete system of Newtonian
equations for the static elastic shell in its self-gravitating
field.

\section{Main Theorem}

The resulting equations of the latter section --- \eqref{Poisson
equation}, \eqref{load equation}, \eqref{boundary Newton} --- are
too complicated to be solved explicitly for general elastic
materials. Therefore, we will resort to analytical methods in the
sequel. In order to use the machinery of the implicit function
theorem we write our field equation and boundary conditions as a map
between Sobolev spaces. Integrating the Poisson equation
\eqref{Poisson equation} and inserting it into the load equation
\eqref{load equation} we find the resulting Newtonian field equation
to be an integro-differential equation of the following form:
\begin{equation*}
\frac{2}{F}\big((T_{r}{}^{r})_{newt}-(T_{\vartheta}{}^{\vartheta})_{newt}\big)
+\big(\frac{dF}{dR}\big)^{-1}(\pd{}{R}T_{r}{}^{r})_{newt}=-\frac{4\pi\mathcal{G}R^2\rho_0^2}{F^4}
\big(\frac{dF}{dR}\big)^{-1}\int_{R_{i}}^{R} F^{2}(\bar{R})d\bar{R}.
\end{equation*}
\noindent The latter together with the boundary condition
\begin{equation*}
\frac{F^{2}}{R^{2}}(T_{r}{}^{r})_{newt}\mid_{\partial{\Bbb{B}}}=0
\end{equation*}
\noindent leads to modelling the characteristic mapping,
$\mathcal{F}$, corresponding to the static elastic shell in the
context of the Newtonian material description. It is obtained from
the system of field equation and boundary conditions in the
Newtonian theory of gravity, that is

\begin{definition} \textbf{S}(tatic)\textbf{E}(lastic)\textbf{S}(hell) \textbf{map}:

\begin{eqnarray*}
&&\mathcal{F}: W^{2,2}\big((R_{i},R_{o}) \times \Bbb{R}\big)
\rightarrow
W^{0,2}\big((R_{i},R_{o}) \times \Bbb{R}\big) \times W^{\frac{1}{2},2}\big(\{ R_{i},R_{o}\} \times \Bbb{R}\big)\\
&&[F(R),\mathcal{G}] \mapsto
\mathcal{F}[F(R),\mathcal{G}]=\big(\hat{E}[F(R)]-\mathcal{G}\hat{e}[F(R)],\hat{b}[F(R)]\big),
\end{eqnarray*}
\noindent where

\begin{eqnarray*}
&&\textbf{Static Elasticity Operator}\\
&&\hat{E}[F(R)]:=\big[2\frac{d^2F}{dR^2}+\frac{4\frac{dF}{dR}}{R}-\frac{4F}{R^{2}}\big]\pd{w}{J_{1}}+
2\frac{dF}{dR}\frac{\partial^{2}{w}}{\partial{J_{1}}\partial{R}}+\big[\frac{4\frac{d^2F}{dR^2}F^{2}}{R^{2}}
+\frac{4\big(\frac{dF}{dR}\big)^{2}F}{R^{2}}-\\
&&\phantom{\hat{E}[F(R)]:=}\frac{4F^{3}}{R^{4}}\big]\pd{w}{J_{2}}+
\frac{4\frac{dF}{dR}F^{2}}{R^{2}}\frac{\partial^{2}{w}}{\partial{J_{2}}\partial{R}}+
\big[\frac{2\frac{d^2F}{dR^2}F^{4}}{R^{4}}+\frac{4\big(\frac{dF}{dR}\big)^{2}F^{3}}{R^{4}}-\\
&&\phantom{\hat{E}[F(R)]:=}\frac{4\frac{dF}{dR}F^{4}}{R^{5}}\big]\pd{w}{J_{3}}+\frac{2\frac{dF}{dR}F^{4}}
{R^{4}}\frac{\partial^{2}{w}}{\partial{J_{3}}\partial{R}},\\
&&\textbf{Force Operator}\\
&&\hat{e}[F(R)]:=\frac{4\pi \rho_{0}^{2}}{F^{2}}\int _{R_{i}}^{R} F^{2}(\bar{R}) d\bar{R},\\
&&\textbf{Boundary Operator}\\
&&\hat{b}[F(R)]:=2\frac{dF}{dR}[\pd{w}{J_{1}}+2\frac{F^{2}}{R^{2}}\pd{w}{J_{2}}+\frac{F^{4}}{R^{4}}
\pd{w}{J_{3}}]\mid_{\partial{\Bbb{B}}}.
\end{eqnarray*}
\end{definition}
\noindent In the above definition $W^{p,k}$ stands for the standard
Sobolev spaces of functions having $p$ weak derivatives in $L^k$ ---
see, for example, \cite{AmbPro;93}.
\\ \textbf{Remark.} The Sobolev
spaces were chosen out of technical reasons. The first factor in the
range of $\mathcal{F}$ rises from considerations concerning the
field equation, the second factor in the range from considerations
concerning the boundary conditions. The weight associated with the
weak derivative in the first factor of the range comes from the
appearance of second derivatives in the static elasticity operator
while that of the second factor is dictated by a lemma in
\cite{Val;88}. We will make use of this lemma later. \\ \\
Before giving our main result we will recall one of our crucial
assumptions and we will consider some technical lemmata that will be
used to prove our main theorem.
\begin{assumption}
We require our system to satisfy pointwise stability. For isotropic
linear elasticity pointwise stability --- see, for example
\cite{MarHug;83} --- is fulfilled when
\begin{equation*}
\mu >0 \quad \ \ \ 3\kappa=3\lambda +2\mu >0,
\end{equation*}
where $\lambda$ and $\mu$ denote the Lam\'e constants.
\end{assumption}
\noindent \textbf{Remark.} The constant $\kappa$ is called the
\emph{bulk modulus}. The inequality $3\kappa>0$ allows negative
$\lambda$ with $ -\frac{2}{3} \mu$ as a lower bound (\emph{auxetic
materials}).
\begin{lemma}
$\mathcal{F} \in C^{1}\bigg(U, W^{0,2}\big((R_{i},R_{o}) \times
\Bbb{R})\big) \times W^{\frac{1}{2},2}\big(\{ R_{i},R_{o}\} \times
\Bbb{R}\big) \bigg)$, where $U$ is an open subset of the Sobolev
space $W^{2,2}((R_{i},R_{o}) \times \Bbb{R})$.
\end{lemma} \noindent
\textbf{Proof.} To check if the SES map is $C^1$, we consider the
special case of $m=0$ and $p=2$ in the assumptions made in a lemma
in \cite{Val;88}. We find that one easily derives the validity of
the assertion for the pair of the following operators, namely, the
static elasticity operator and the boundary operator. To show that
the force operator is $C^1$, we make use of a corollary in
\cite{AbrMarRat;88} that reads as follows\begin{corollary} If
$\mathcal{F}:U \subset W^{2,2}\big((R_{i},R_{o}) \times \Bbb{R}\big)
\rightarrow W^{0,2}\big((R_{i},R_{o}) \times \Bbb{R}\big) \times
W^{\frac{1}{2},2}\big(\{ R_{i},R_{o}\}
 \times \Bbb{R}\big)$ is $C^1$-G\^{a}teaux then it is $C^1$ and the two
derivatives coincide.
\end{corollary}
\noindent Essentially, the force operator corresponds to
$\frac{\partial}{\partial F}U[F(R)]$. The computation of the
G\^{a}teaux derivative of $\frac{\partial}{\partial F}U[F(R)]$ gives
us
\begin{equation*}
\frac{d}{d\tau}(\frac{\partial}{\partial
F}U[F_{(\tau)}(R)])_{\tau=0}=D(\frac{\partial}{\partial
F}U[F_{(0)}(R)])\cdot \chi,
\end{equation*}
where
\begin{eqnarray*}
&&F_{(\tau)}(R)=R+\tau\chi(R); \ \ \ F_{0}(R)=R\\
&&\chi(R) \in W^{2,2}((R_{i},R_{o})).
\end{eqnarray*}
\noindent The G\^{a}teaux derivative of $\frac{\partial}{\partial
F}U[F(R)]$ reads
\begin{equation*}
\frac{d}{d\tau}(\frac{\partial}{\partial
F}U[F_{(\tau)}(R)])_{\tau=0}=\frac{4\pi \mathcal{G}
\rho_{0}}{3}\big(1+2\frac{R_{i}^{3}}{R^{3}}\big)\chi=D(\frac{\partial}{\partial
F}U[F_{(0)}(R)])\cdot \chi.
\end{equation*}
\noindent It is an element of the space of linear continous maps
$\mathcal{L}(W^{2,2}((R_{i},R_{o})
\times \Bbb{R}), W^{0,2}((R_{i},R_{o}) \times \Bbb{R}))$. \\
Now, $R \mapsto D(\frac{\partial}{\partial F}U[F_{(0)}])$ is
continuous for $R \in [R_{i},R_{o}]$ since the linear operator
defined by the directional
derivative is clearly bounded.\\ \\
Thus, the force operator is $C^{1}$-G\^{a}teaux, from where it
follows that the force operator $\hat{e}[F(R)]$ is $C^{1}$.
Therefore, we derive the SES map $\mathcal{F}[F(R),\mathcal{G}]$
--- as defined above --- is $C^{1}$. $\Box$
\begin{lemma}
$\mathcal{F}[F_{0}=R,\mathcal{G}_{0}=0] \equiv (0,0)$.
\end{lemma}
\noindent \textbf{Proof.} For
$\mathcal{F}[F=F_{0},\mathcal{G}=\mathcal{G}_{0}]\equiv (0,0)$ we
find $F_{0}$ and $\mathcal{G}_{0}$ to be such that  $F_{0}=R$ and
$\mathcal{G}_{0}=0$ as --- inter alia --- the invariants of
$(H^{A}{}_{B})_{newt}$ on which $w$ depends are constant for
$F_{o}=R$ and $\mathcal{G}_{0}=0$. We have
\begin{equation*}
\mathcal{F}[F_{0}=R,\mathcal{G}_{0}=0]\equiv 0. \ \ \ \Box
\end{equation*}  \noindent
\begin{lemma}
$D_{F}\mathcal{F}[F_{0},\mathcal{G}_{0}]$ is an isomorphism from
$W^{2,2}((R_{i},R_{o}) \times \Bbb{R})$ onto
$W^{0,2}\big((R_{i},R_{o}) \times \Bbb{R})\big) \times
W^{\frac{1}{2},2}\big(\{ R_{i},R_{o}\} \times \Bbb{R}\big)$.
\end{lemma}
\noindent \textbf{Proof.} We divide the proof of this lemma in three
parts:\\
 \textbf{First task.} From the Taylor
expansion of $\mathcal{F}$
--- which is supposed to be at least $C^{r+1}$ --- around $R$ for
$\delta F$ sufficiently small we see that\begin{center}
$\mathcal{F}[F_{0}+\delta
F,\mathcal{G}_{0}]=D_{F}\mathcal{F}(F_{0}=R,\mathcal{G}_{0}=0)\delta
F+o(||\delta F||^{2}).$
\end{center}
\noindent The components of the linearised SES map read

\begin{eqnarray*}
&&\stackrel{lin}{\hat{E}}[\delta F]=\bigg(\frac{d^2}{dR^2}(\delta F)+\frac{2}{R}\frac{d}{dR}(\delta F)-\frac{2(\delta F)}{R^{2}}\bigg)(\lambda +2\mu),\\
&&\stackrel{lin}{\hat{e}}[\delta F]=\frac{4\pi \rho_{0}^{2}}{3R^{2}}(R^{3}-R_{i}^{3}),\\
&&\stackrel{lin}{\hat{b}}[\delta F]=\left[\frac{d}{dR}(\delta
F)(\lambda+2\mu)+\frac{2}{R}\lambda (\delta
F)\right]\mid_{\partial{\Bbb{B}}},
\end{eqnarray*}
\begin{equation*}
\text{hence} \stackrel{lin}{\mathcal{F}}[\delta
F,\mathcal{G}]:=(\stackrel{lin}{\hat{E}}[\delta F]-
\mathcal{G}\stackrel{lin}{\hat{e}}[\delta F],\stackrel{lin}{\hat{b}}[\delta F]).\\
\end{equation*}
\noindent Neglecting higher orders, we are now able to concentrate
on $\mathcal{F}[F_{0}+\delta F,\mathcal{G}_{0}]$ to see if
$D\mathcal{F}(F_{0}=R,\mathcal{G}_{0}=0)$ is an isomorphism.\\
 \textbf{Second task.}
Next, we will show that $\mathcal{F}[F_{0}+\delta
F,\mathcal{G}_{0}]$ is injective.\\ \\ We compute the solution of
$\stackrel{lin}{\hat{E}}[\delta F]=0$. The latter reads
\begin{equation*}
R^{2}\frac{d^2}{dR^2}(\delta F)+2R\frac{d}{dR}(\delta F)-2(\delta
F)=0,
\end{equation*}
\noindent which is an ordinary differential equation of the Eulerian
type.

The solution of the above equation reads,
\begin{equation*}
\delta F(R)=u_{1}R+u_{2}R^{-2},
\end{equation*}
\noindent where $u_{k} \in \Bbb{R}$ for $k=1,2$. Inserting the
solution $\delta F$ in the linearised boundary conditions

\begin{equation*}
\bigg(\frac{d}{dR}(\delta F)(\lambda +2\mu)+\frac{2(\delta F)}{R}
\lambda \bigg)|_{\partial \Bbb{B}}=0,
\end{equation*}
\noindent we derive

\begin{equation*}
\bigg(u_{1}(3\lambda+2\mu)-u_{2}\frac{4\mu}{R^{3}}\bigg)|_{\partial
\Bbb{B}}=0.
\end{equation*}
\noindent We can write the latter equations in matricial form as
\begin{equation}
\label{system} \Bbb{A}\vec{u}=0,
\end{equation}
\noindent where
\begin{equation*}
\Bbb{A}=\begin{pmatrix}
3\lambda+2\mu & -4\mu\frac{1}{R^{3}_{i}}\\
3\lambda+2\mu & -4\mu\frac{1}{R^{3}_{o}}\\
\end{pmatrix},
\quad \vec{u}=\begin{pmatrix}
u_{1}\\
u_{2}\\
\end{pmatrix}.
\end{equation*}
\noindent The determinant of the coefficients matrix of this system
of linear equations, $\det{\Bbb{A}}$, is
\begin{equation*}
\det{\Bbb{A}}=12\kappa\mu\big(\frac{1}{R^{3}_{i}}-\frac{1}{R^{3}_{o}}\big),
\end{equation*}
\noindent
where $3\kappa=3\lambda+2\mu$.\\
Assume for the moment that the above system \eqref{system} has
solutions other than the trivial solution, that is equivalent to the
fact that $\det{\Bbb{A}}=0$. If this is the case it follows from
pointwise stability that $R_{i}=R_{o}$. However, $R_{i}=R_{o}$ is
not an allowed solution. Thus, $\det{\Bbb{A}}\neq 0$ and the system
\eqref{system} has only the trivial solution.
 Therefore, $\stackrel{lin}{\mathcal{F}}[\delta
F]=0$ has no kernel for $\mathcal{G}=0$ except for the trivial
solution. That is, the mapping $\stackrel{lin}{\mathcal{F}}[\delta
F]$ is injective for $\mathcal{G}=0$.\\ \\
\bigskip
\textbf{Third task.} Now, we will show that
$\stackrel{lin}{\mathcal{F}}[\delta F]$ is surjective for
$\mathcal{G}=0$. \\  Let us write
\begin{equation}
\label{surjective} \frac{d^2}{dR^2}(\delta
F)+\frac{2}{R}\frac{d}{dR}(\delta F)-2\frac{\delta
F}{R^{2}}=\phi(R),
\end{equation}
\noindent and suppose that
\begin{equation*}
\phi(R) \in W^{0,2} \big((R_{i},R_{o})\big)\times
W^{\frac{1}{2},2}\big(\{R_{i},R_{o}\}\big).
\end{equation*}
\noindent The solution $(\delta F)_{g}(R)$ of the above differential
equation, \eqref{surjective}, consists of the sum over the solution
$(\delta F)_{h}(R)$ of the homogenous differential equation, that is
$\phi(R)=0$, and a particular solution $(\delta F)_{p}(R)$ of the
inhomogenous equation. The solution to the homogenous case as
already  computed before is
\begin{equation*}
(\delta F)_{h}(R)=u_{1}R+u_{2}R^{-2}.
\end{equation*}
\noindent One way of obtaining a particular solution to
\eqref{surjective} is by means of the following ansatz
\begin{equation*}
(\delta F)_{p}(R)=v(R)R.
\end{equation*}

\noindent After a few calculations we get
\begin{equation}
\label{v}
v(R)=\int_{R_{i}}^{R}\frac{1}{\bar{R}^{4}}\bigg(\int_{R_{i}}^{\bar{R}}
\stackrel{=}{R}^{3}\phi(\stackrel{=}{R})\
d\stackrel{=}{R}+R_{i}^{4}\frac{d}{dR}v(R_{i})\bigg)d\bar{R}.
\end{equation}
\noindent Since we are searching for any particular solution, it is
possible to choose the constants of integration in a way that the
last term in the above equation vanishes. Therefore, the solution of
equation \eqref{surjective} reads
\begin{equation}
\label{solution} (\delta
F)_{g}(R)=u_{1}R+u_{2}R^{-2}+R\int_{R_{i}}^{R}\frac{1}{\bar{R}^{4}}\bigg(\int_{R_{i}}^{\bar{R}}
\stackrel{=}{R}^{3}\phi(\stackrel{=}{R})\
d\stackrel{=}{R}\bigg)d\bar{R}.
\end{equation}
\noindent Inserting this solution into the boundary conditions
\begin{equation*}
\bigg(\frac{d}{dR}(\delta F)_{g}(R)+\frac{\gamma}{R}(\delta
F)_{g}(R)\bigg)|_{\partial{\Bbb{B}}}=0,
\end{equation*}
where $\gamma=2\lambda(\lambda +2 \mu)^{-1}$, we can determine the
form of the constants $u_{1}$ and $u_{2}$ in terms of $v(R_{i})$ and
$v(R_{o})$:

\begin{eqnarray}
\nonumber
&&u_{2}=-\frac{R_{i}^{3}R_{o}^{3}}{(\gamma-2)(R_{o}^{3}-R_{i}^{3})}\bigg(
R_{i}\frac{d}{dR}v(R_{i})-R_{o}\frac{d}{dR}v(R_{o})+(1+\gamma)\big(v(R_{i})-v(R_{o})\big)\bigg),\\
\label{u constants}
&&u_{1}=-\frac{1}{1+\gamma}\bigg(u_{2}(\gamma-2)R_{i}^{-3}+R_{i}\frac{d}{dR}v(R_{i})
+\gamma v(R_{i})\bigg).
\end{eqnarray}
\noindent Let us write
\begin{equation*}
(\delta F)_{g}(R)=(\delta F)_{g_1}(R) +(\delta F)_{g_2}(R) +(\delta
F)_{g_3}(R),
\end{equation*}
\noindent where
\begin{eqnarray*}
&&(\delta F)_{g_1}(R)=u_{1}\big(\phi(R_{i}), \phi(R_{o})\big)R, \\
&&(\delta F)_{g_2}(R)=u_{2}\big(\phi(R_{i}),
\phi(R_{o})\big)R^{-2},\\
&&(\delta F)_{g_3}(R)=v\big(\phi(R)\big)R.
\end{eqnarray*}
\noindent Using the fundamental theorem of calculus it is easy to
see that
\begin{equation*}
(\delta F)_{g_1}(R) \in W^{2,2}\big((R_{i},R_{o})\big), \quad
\text{and} \ \ \ (\delta F)_{g_2}(R) \in
W^{2,2}\big((R_{i},R_{o})\big).
\end{equation*}
\noindent Concerning $(\delta F)_{g_3}(R)$, the following is true
\begin{eqnarray*}
&&\frac{d}{dR}(\delta F)_{g_3}(R) \in C^{0}\big((R_{i},R_{o})\big),\\
&&\bigg(\int_{R_{i}}^{R_{o}}|\frac{d}{dR}(\delta
F)_{g_3}(R)|^{2}dR\bigg)^{\frac{1}{2}}<\infty.
\end{eqnarray*}
\noindent The same arguments for the second weak derivative of
$(\delta F)_{g_3}(R)$ lead to the result that
\begin{equation*}
(\delta F)_{g_3}(R) \in W^{2,2}\big((R_{i},R_{o})\big).
\end{equation*}
\noindent We conclude that the solution $(\delta F)_{g}(R)$ of the
inhomogenous differential equation \eqref{surjective} with $\phi(R)
\in W^{0,2} \big((R_{i},R_{o})\big)\times
W^{\frac{1}{2},2}\big(\{R_{i},R_{o}\}\big)$ is an element of the
Sobolev space as required, namely
\begin{equation*}
(\delta F)_{g}(R) \in W^{2,2}\big((R_{i},R_{o})\big).
\end{equation*}
\noindent Hence, $\stackrel{lin}{\mathcal{F}}[\delta F]$ is
surjective for $\mathcal{G}=0$. \\ \\
\textbf{Remark.} At this point
let us say something about the possibility of taking the limit $R_i
\rightarrow 0$, which yields a static elastic sphere. It can be
checked that
\begin{equation*}
\phi=O(R^{-p}),
\end{equation*}
\noindent where $p<1$, $\phi \in W^{0,2}\big((R_i,R_o)\big) \times
W^{\frac{1}{2},2}\big(\{R_i,R_o\}\big)$. Therefore we conclude from
\eqref{v}  that
\begin{eqnarray*}
&&v(R)=\int_{R_i}^{R}\bar{R}^{-4}\bigg(\int_{R_i}^{\bar{R}}\stackrel{=}{R}^{3-p}d\stackrel{=}{R}\bigg)\\
&&\phantom{v(R)}=
\frac{1}{4-p}\bigg(\frac{R^{1-p}}{1-p}+\frac{R_i^{4-p}}{3R^3}-\frac{R_i^{1-p}}{1-p}-\frac{R_i^{1-p}}{3}\bigg)=O(R_i^0),
\end{eqnarray*}
\noindent and
\begin{equation*}
v(R_i)=O(R_i^q), \quad \ \ \ R_iv(R_i)=O(R_i^q),
\end{equation*}
\noindent where $q>0$. Thus, it follows from \eqref{u constants}
that
\begin{equation*}
u_2\longrightarrow 0 \quad \text{as\ }R_i\rightarrow 0,
\end{equation*}
\noindent and the constant $u_1$ is bounded for $R_i \rightarrow 0$.
The solution of the linearised system of equations for the static
elastic sphere reads
\begin{equation*}
(\delta F)(R)\stackrel{R_i\rightarrow 0}{=}u_1(R_i\rightarrow
0)R+R\int_0^R
\bar{R}^{-4}\bigg(\int_0^{\bar{R}}\stackrel{=}{R}^{3}\phi(\stackrel{=}{R})\bigg)d\bar{R}.
\end{equation*}

\noindent Again, the linearised mapping
$\stackrel{lin}{\mathcal{F}}[\delta F]$ for $\mathcal{G}=0$ is both
injective and surjective. Therefore,
\begin{equation*}
D_{F}\mathcal{F}(F_{0},\mathcal{G}_{0}): W^{2,2}((R_{i},R_{o}),
\Bbb{R}) \rightarrow W^{0,2}\big((R_{i},R_{o}),\Bbb{R})\big) \times
W^{\frac{1}{2},2}\big(\{ R_{i},R_{o}\} ,\Bbb{R}\big)
\end{equation*}
\noindent is an isomorphism. $\Box$\\ \\ Now we are ready to state
and prove our main result.
\begin{main}
Let \begin{eqnarray*} &&\mathcal{F}: W^{2,2}\big((R_{i},R_{o})
\times \Bbb{R}\big) \rightarrow
W^{0,2}\big((R_{i},R_{o}) \times \Bbb{R}\big) \times W^{\frac{1}{2},2}\big(\{ R_{i},R_{o}\} \times \Bbb{R}\big)\\
&&[F(R),\mathcal{G}] \mapsto
\mathcal{F}[F(R),\mathcal{G}]=\big(\hat{E}[F(R)]-\mathcal{G}\hat{e}[F(R)],\hat{b}[F(R)]\big),
\end{eqnarray*}
\noindent where

\begin{eqnarray*}
&&\hat{E}[F(R)]:=\big[2\frac{d^2F}{dR^2}+\frac{4\frac{dF}{dR}}{R}-\frac{4F}{R^{2}}\big]\pd{w}{J_{1}}+2
\frac{dF}{dR}\frac{\partial^{2}{w}}{\partial{J_{1}}\partial{R}}+\big[\frac{4\frac{d^2F}{dR^2}F^{2}}{R^{2}}
+\frac{4\big(\frac{dF}{dR}\big)^{2}F}{R^{2}}-\\
&&\phantom{\hat{E}[F(R)]:=}\frac{4F^{3}}{R^{4}}\big]\pd{w}{J_{2}}+
\frac{4\frac{dF}{dR}F^{2}}{R^{2}}\frac{\partial^{2}{w}}{\partial{J_{2}}\partial{R}}+\big[
\frac{2\frac{d^2F}{dR^2}F^{4}}{R^{4}}+\frac{4\big(\frac{dF}{dR}\big)^{2}F^{3}}{R^{4}}-\\
&&\phantom{\hat{E}[F(R)]:=}\frac{4\frac{dF}{dR}F^{4}}{R^{5}}\big]\pd{w}{J_{3}}+
\frac{2\frac{dF}{dR}F^{4}}{R^{4}}\frac{\partial^{2}{w}}{\partial{J_{3}}\partial{R}},\\
&&\hat{e}[F(R)]:=\frac{4\pi \rho_{0}^{2}}{F^{2}}\int _{R_{i}}^{R} F^{2}(\bar{R}) d\bar{R},\\
&&\hat{b}[F(R)]:=2\frac{dF}{dR}[\pd{w}{J_{1}}+2\frac{F^{2}}{R^{2}}\pd{w}{J_{2}}+\frac{F^{4}}{R^{4}}
\pd{w}{J_{3}}]\mid_{\partial{\Bbb{B}}}.
\end{eqnarray*}
Then there exists a neighbourhood $N_{\mathcal{G}}$ of
$\mathcal{G}_{0}$, $N_{\mathcal{G}} \subset \Bbb{R}$ and a
neighbourhood $N_{F}$ of $F_{0}$, $N_{F} \subset
W^{2,2}((R_{i},R_{o}) \times \Bbb{R})$ and a map
 $\bar{F} \in C^{1}(N_{\mathcal{G}},N_{F})$ such that
\begin{itemize}
\item[(i)] $\mathcal{F}[\bar{F}(R,\mathcal{G}),\mathcal{G}]=(0,0)\ \ \   \forall \mathcal{G} \in N_{\mathcal{G}}$,\\
\item[(ii)]$\mathcal{F}[F(R),\mathcal{G}]=(0,0), \ \ \
[F(R),\mathcal{G}] \in N_{\mathcal{G}} \times N_{F}$, implies
$F(R)=\bar{F}(R,\mathcal{G})$,\\
\item[(iii)]
$\bar{F}'(R,\mathcal{G})=-\big(D_{F}\mathcal{F}[\bar{F}(R,\mathcal{G}),\mathcal{G}]\big)^{-1}
\circ D_\mathcal{G}\mathcal{F}[\bar{F}(R,\mathcal{G}),\mathcal{G}]$,
where $\mathcal{G} \in N_{\mathcal{G}}$.
\end{itemize}
\end{main}
\noindent \textbf{Proof.} The proof of the above theorem is a direct
consequence from Lemma 1, Lemma 2 and Lemma 3 as well as from the
implicit function theorem --- see \cite{AmbPro;93}. $\Box$

\section{Concluding remarks}
From the above theorem it is now clear that as long as pointwise
stability is fulfilled, for example perfect fluids will not satisfy
this condition, stable states of the static elastic shell in its own
gravitational field
exist if the body is sufficiently small.\\
The present work can be extended in several directions. It would be
of great interest to provide the analogous analytic proof for the
existence of stable states in the theory of Einsteinian gravity.
Furthermore, one could undertake some numerical investigations
picking out some specific realistic material. It would also be
possible to replace the vacuum inside the hollow centre of the shell
by some matter, for example air.

\section*{Acknowledgments}
The author would like to thank Prof.Robert Beig for his advice and
suggestions while this project was being carried, Prof.Bernd Schmidt
who recommended considering this problem, Dr.Juan A. Valiente Kroon
for a critical reading of the manuscript, and Dr.Michael
Wernig-Pichler for valuable discussions.

\appendix
\section{Solution to the system of linearised equations}

The Newtonian system of field equation and boundary conditions is
too complicated to be solved for general material, still the
linearised equations can easily be solved analytically. In addition,
we will list a few examples for which we investigate the behaviour
of the static elastic shell or the static elastic sphere as a
special case
 taking into consideration various models of matter.
 We consider the system of linearised equations

\begin{eqnarray*}
&&\frac{d^2}{dR^2}(\delta F)+\frac{2}{R}\frac{d}{dR}(\delta F)-2\frac{(\delta F)}{R^{2}}=\phi(R),\\
&&\bigg(\frac{d}{dR}(\delta F)+\frac{\gamma}{R}(\delta
F)\bigg)_{\partial{\Bbb{B}}}=0,
\end{eqnarray*}
\noindent where
\begin{equation*}
\phi(R)=\mathcal{K}\big(R-\frac{R_{i}^{3}}{R^{2}}\big), \quad
\mathcal{K}=\mathcal{G}\frac{4\pi\rho_{0}^{2}}{3(\lambda+2\mu)},
\quad \text{and} \ \ \ \gamma=\frac{2\lambda}{\lambda+2\mu}.
\end{equation*}
\noindent As we know from section 4, the solution to the above
system for a general $\phi(R)$
--- see equations \eqref{solution}, \eqref{v} and \eqref{u constants}--- reads
\begin{equation}
\label{solG} (\delta
F)(R)=\bigg(u_{1}\big(\phi(R_{i}),\phi(R_{o})\big)+v\big(\phi(R)\big)\bigg)R+
u_{2}\big(\phi(R_{i}),\phi(R_{o})\big)R^{-2},
\end{equation}
\noindent where
\begin{eqnarray}
\label{vG}
&&v(R)=\int_{R_{i}}^{R}\frac{1}{\bar{R}^{4}}\bigg(\int_{R_{i}}^{\bar{R}}
\stackrel{=}{R}^{3}\phi(\stackrel{=}{R})\ d\stackrel{=}{R}\bigg)d\bar{R},\\
\nonumber
&&u_{2}=-\frac{R_{i}^{3}R_{o}^{3}}{(\gamma-2)(R_{o}^{3}-R_{i}^{3})}\bigg(
R_{i}\frac{d}{dR}v(R_{i})-R_{o}\frac{d}{dR}v(R_{o})+(1+\gamma)\big(v(R_{i})-v(R_{o})\big)\bigg),\\
\label{u1G}
&&\\
\label{u2G}
&&u_{1}=-\frac{1}{1+\gamma}\bigg(u_{2}(\gamma-2)R_{i}^{-3}+R_{i}\frac{d}{dR}v(R_{i})
+\gamma v(R_{i})\bigg).
\end{eqnarray}
\noindent Inserting
$\phi(R)=\mathcal{K}\big(R-\frac{R_{i}^{3}}{R^{2}}\big)$ into
\eqref{vG} we obtain
\begin{equation*}
v(R)=\mathcal{K}\bigg(\frac{R^{2}}{10}+\frac{R_{i}^{3}}{2R}-\frac{R_{i}^{5}}{10R^{3}}-\frac{R_{i}^{2}}{2}\bigg).
\end{equation*}
\noindent And the following holds,
\begin{eqnarray*}
&& v(R_{i})=\frac{d}{dR}v(R_{i})=0,\\
&& v(R_{o})=\mathcal{K}\bigg(\frac{R_{o}^{2}}{10}+\frac{R_{i}^{3}}{2R_{o}}-\frac{R_{i}^{5}}{10R_{o}^{3}}-\frac{R_{i}^{2}}{2}\bigg),\\
&&
\frac{d}{dR}v(R_{o})=\mathcal{K}\bigg(\frac{R_{o}}{5}-\frac{R_{i}^{3}}{2R_{o}^{2}}+\frac{3R_{i}^{5}}{10R_{o}^{4}}\bigg).
\end{eqnarray*}
\noindent With these results it is clear that
\begin{eqnarray*}
&&u_{2}=\mathcal{K}\frac{R_{i}^{3}R_{o}^{3}}{(\gamma-2)(R_{o}^{3}-R_{i}^{3})}
\bigg(\frac{3+\gamma}{10}R_{o}^{2}-\frac{\gamma-2}{10}\frac{R_{i}^{5}}{R_{o}^{3}}-
\frac{1+\gamma}{2}R_{i}^{2}+\frac{\gamma}{2}\frac{R_{i}^{3}}{R_{o}}\bigg),\\
&&u_{1}=-\mathcal{K}\frac{R_{o}^{3}}{(1+\gamma)(R_{o}^{3}-R_{i}^{3})}
\bigg(\frac{3+\gamma}{10}R_{o}^{2}-\frac{\gamma-2}{10}\frac{R_{i}^{5}}{R_{o}^{3}}-
\frac{1+\gamma}{2}R_{i}^{2}+\frac{\gamma}{2}\frac{R_{i}^{3}}{R_{o}}\bigg).
\end{eqnarray*}
\noindent After some simplifying computations  we have, in the end,
the solution to the system of linearised equations, namely
\begin{center}
$(\delta F)(R)=\mathcal{G}\frac{4\pi \rho_{0}^{2}}{3(\lambda+2\mu)}
\bigg(\frac{R^{3}}{10}+\bar{u}_{1}R+\bar{u}_{2}R^{-2}+\frac{R_{i}^{3}}{2}\bigg)$,
\end{center}
\noindent where
\begin{eqnarray*}
&&\bar{u}_{1}=-\frac{R_{o}^{3}}{3\kappa(R_{o}^{3}-R_{i}^{3})}
\big(\frac{5\lambda
+6\mu}{10}R_{o}^{2}-\frac{3(5\lambda+2\mu)}{10}\frac{R_{i}^{5}}{R_{o}^{3}}
+\lambda \frac{R_{i}^{3}}{R_{o}}\big),\\
&&
\bar{u}_{2}=\frac{R_{i}^{3}R_{o}^{3}}{4\mu(R_{o}^{3}-R_{i}^{3})}\big(\frac{5\lambda
+6\mu}{10}R_{o}^{2} -\frac{3(5\lambda+2\mu)}{10}R_{i}^{2}+\lambda
\frac{R_{i}^{3}}{R_{o}}\big),
\end{eqnarray*}
\noindent and $3\kappa=3\lambda +2\mu$.
Note that for $R_i=R_o$, $\bar{u}_1=\bar{u}_2=0$.\\ \\
Next, we want to give some common examples of elastic materials and
investigate in which regions the material is stretched and where it
is compressed. The (R,R) component of the strain tensor
$(H^{A}{}_{B})$ measures the compression, $(H^{R}{}_{R})<0$, as well
as the stretching, $(H^{R}{}_{R})>0$, of the material. For the
linearised system of equations of the static elastic shell, we have
\begin{equation}
\label{strain} (\stackrel{lin}{H}^{R}{}_{R})(R)=\frac{d}{dR}(\delta
F)(R)=\frac{\mathcal{G}4\pi
\rho_{0}^2}{3(\lambda+2\mu)}\big(\frac{3R^2}{10}+\bar{u}_1-2\bar{u}_2R^{-3}\big).
\end{equation}
\noindent \textbf{Example 1: Ideal cork.} We consider a material
where the Lam\'{e} constant $\lambda=0$. This is a model for ideal
cork\footnote{See, for example, the following homepage: {\tt
http://home.att.net/ $\sim$ numericana/answer/physics.htm.}}.
 For such a material
\begin{equation*}
\bar{u}_1(\lambda=0)=-\frac{3}{10}R_o^2\frac{1-b^5}{1-b^3}<0, \quad
\ \ \
\bar{u}_2(\lambda=0)=\frac{3}{20}R_o^5\frac{b^3(1-b^2)}{1-b^3}>0,
\end{equation*}
\noindent where $R_i=bR_o$, $b\in[0,1)$. If $b=0$ we have the
special case of a static elastic sphere --- see example 3. We apply
D\'{e}scartes' rule of signs\footnote{D\'{e}scartes' rule of signs
states that for a given polynomial the number of sign changes of the
coeffients of the polynomial gives the maximum number of positive
roots --- see e.g. \cite{RahSch;02}.} to the following polynomial:
\begin{equation*}
R^5+\frac{10}{3}\bar{u}_1R^3-\frac{20}{3}\bar{u}_2=0,
\end{equation*}\noindent
and see that in this case we have at most one positive zero. Next,
we consider $(\stackrel{lin}{H}^{R}{}_{R})(R_i)$ and
$(\stackrel{lin}{H}^{R}{}_{R})(R_o)$ and investigate their signs:
\begin{equation*}
(\stackrel{lin}{H}^{R}{}_{R})(R_i)\stackrel{\lambda=0}{=}\frac{2}{5}\frac{\mathcal{G}\pi
  \rho_0^2R_o^2}{\mu}\frac{b^2-1}{1-b^3}<0, \quad \ \ \
(\stackrel{lin}{H}^{R}{}_{R})(R_o)\stackrel{\lambda=0}{=}\frac{2}{5}\frac{\mathcal{G}\pi
  \rho_0^2R_o^2}{\mu}\frac{b^3(b^2-1)}{1-b^3}<0.
\end{equation*}
\noindent Thus, we know that for this particular material, there
exists no positive zeros and the material is compressed in the whole
shell. Alternatively to the latter discussion, we can investigate
what happens to the solution $(\delta F)(R)$ at the inner radius,
$R_i$, and at the outer radius $R_o$:
\begin{eqnarray*}
&&(\delta F)(R_i)\stackrel{\lambda=0}{=}
\frac{\mathcal{G}2\pi\rho_0^2}{3\mu}
\big(\frac{3R_i^3}{5}+\bar{u}_1(\lambda=0)R_{i}+\bar{u}_2(\lambda=0)R_i^{-2}\big)\\
&&\phantom{(\delta F)(R_i)}=\frac{\mathcal{G}\pi\rho_0^2R_o^3b}{10(1-b^3)}(-2b^5+3b^2-1)>0,\\
&&(\delta F)(R_o)\stackrel{\lambda=0}{=}
\frac{\mathcal{G}2\pi\rho_0^2}{3\mu}
\big(\frac{R_o^3+5R_i^3}{10}+\bar{u}_1(\lambda=0)R_{o}+\bar{u}_2(\lambda=0)R_o^{-2}\big)\\
&&\phantom{(\delta
F)(R_o)}=\frac{\mathcal{G}\pi\rho_0^2R_o^3}{10(1-b^3)}(-10b^6+11b^3+6b^5-3b^2-4)<0.
\end{eqnarray*}
That means that under the influence of the gravitational field of
the shell itself, the inner radius increases and the outer radius
decreases. Therefore, the body is compressed.\\ \\
\textbf{Example 2: Ideal rubber.} Here, we want to investigate the
model for ideal rubber\footnote{See {\tt http://home.att.net/ $\sim$
numericana/answer/physics.htm.}}, where $\mu\ll1$. We have
\begin{eqnarray*}
&&\bar{u}_1(\mu\ll1)=-\frac{R_o^2}{6(1-b^3)}(-3b^5+2b^3+1)<0, \\
&&\bar{u}_2(\mu\ll1)=-\frac{\lambda
R_o^5}{8\mu(1-b^3)}b^3(2b^3-3b^2+1)<0
\text{\ if \ } \lambda <0,\\
&& \bar{u}_2(\mu\ll1)=-\frac{\lambda
R_o^5}{8\mu(1-b^3)}b^3(2b^3-3b^2+1)>0 \text{\ if \ } \lambda >0.
\end{eqnarray*}
\noindent From
\begin{eqnarray*}
&&(\stackrel{lin}{H}^{R}{}_{R})(R_i)\stackrel{\mu\ll1}{\simeq}-\frac{\mathcal{G}\pi\rho_{0}^{2}R_o^2}{3\mu (1-b^3)}(2b^3-3b^2+1)<0, \\
&&(\stackrel{lin}{H}^{R}{}_{R})(R_o)\stackrel{\mu\ll1}{\simeq}-\frac{\mathcal{G}\pi\rho_{0}^{2}R_o^2b^3}{3\mu
(1-b^3)}(2b^3-3b^2+1)<0,
\end{eqnarray*}
we see that for $\lambda>0$ the material is compressed in the whole
shell. \\ \\
\textbf{Example 3: Elastic sphere.} Specialising on the elastic
sphere\footnote{This special case has already been considered in
\cite{Lov;44} and \cite{LanLife;98}. }, we let $R_{i} \rightarrow 0$
and see that
\begin{equation*}
\bar{u}_{1}(R_i\rightarrow 0)=
-\frac{R_{o}^{2}}{10}\big(1+\frac{2c_{2}^{2}}{3\kappa}\big)<0, \quad
\text{and} \ \ \ \bar{u}_{2}(R_i\rightarrow 0)=0,
\end{equation*}
\noindent where $c_{2}^{2}=\lambda +2\mu$. \\
\textbf{Remark.} The constant $c_{2}$ is the speed of the
propagation of the progressive, strongly elliptic wave --- see, for
example, \cite{MarHug;83}.\\
  Example 3 has exactly one positive
zero. In the case of the sphere the solution reduces to
\begin{equation*}
(\delta F)_{R_{i}\rightarrow 0}(R)=\mathcal{G}\frac{4\pi
\rho_{o}^{2}}{30c^{2}_{2}}\bigg(R^{3}-RR_{o}^{2}\big(1+\frac{2}{3}\frac{c_{2}^{2}}{\kappa}\big)\bigg).
\end{equation*}
\noindent Considering the latter equation, we see that within a
sphere of radius
 $\bar{R}=R_{o}\sqrt{\frac{1}{3}\big(1+\frac{2}{3}\frac{c_{2}^{2}}{\kappa}\big)}$ the material is
 compressed since
\begin{equation*}
(\stackrel{lin}{H}^{R}{}_{R})\stackrel{R_{i}\rightarrow 0}{<}0,
\end{equation*}
\noindent whereas outside this sphere the material is stretched:
\begin{equation*}
(\stackrel{lin}{H}^{R}{}_{R})\stackrel{R_{i}\rightarrow 0}{>}0.
\end{equation*}
\noindent Note that if $\lambda=0$ (ideal cork) we have the
following
\begin{equation*}
(\stackrel{lin}{H}^{R}{}_{R})(R_o)\stackrel{R_{i}\rightarrow 0}{=}0.
\end{equation*}

\end{document}